\documentclass{jnmauth}

\usepackage{graphicx}
\usepackage{amssymb}
\usepackage{bm}

\newcommand{\be}{\begin{equation}}
\newcommand{\ee}{\end{equation}}

\newcommand\pictc[5]{\begin{figure}
                   \centerline{
                   \includegraphics*[width=#1\textwidth]{#3}}
               \protect\caption{\protect\label{fig:#4} #5}
                \end{figure}            }

\newcommand\pict[4][0.65]{\pictc{#1}{!tb}{#2}{#3}{#4}}
\newcommand\rpict[1]{\ref{fig:#1}}

\newcommand\leqt[1]{\protect\label{eq:#1}}
\newcommand\reqtn[1]{\ref{eq:#1}}
\newcommand\reqt[1]{(\reqtn{#1})}

\newcommand\lsect[1]{\protect\label{sect:#1}}
\newcommand\rsect[1]{\ref{sect:#1}}

\newcounter{Fig}

\begin{document}
\JNM{1}{6}{00}{28}{00}

\runningheads{A.A. Sukhorukov, I.V. Shadrivov, and Yu.S.
Kivshar}{Wave scattering by metamaterial wedges and interfaces}

\title{Wave scattering by metamaterial wedges and interfaces}

\author{Andrey A. Sukhorukov\corrauth, Ilya V. Shadrivov, and Yuri S. Kivshar}

\address{Nonlinear Physics Centre, Research School of Physical Sciences and Engineering,
Australian National University, Canberra ACT 0200, Australia}

\corraddr{ans124@rsphysse.anu.edu.au}

\noreceived{}
\norevised{}
\noaccepted{}

\begin{abstract}
We outline specific features of numerical simulations of
metamaterial wedges and interfaces. We study the effect of different positioning of a grid in the Yee method, which is necessary to obtain consistent convergence in modeling of interfaces with metamaterials characterized by negative dielectric permittivity and negative magnetic permeability. We demonstrate however that, in the framework of the continuous-medium approximation, wave scattering on the wedge may result in a resonant excitation of surface waves
with infinitely large spatial frequencies, leading to non-convergence of the simulation results that depend on the discretization step.
\end{abstract}


\section{INTRODUCTION}

Recent theoretical \cite{Pendry:1996-4773:PRL,
Pendry:1999-2075:ITMT, Markos:2002-36622:PRE, Markos:2002-33401:PRB} and experimental \cite{Smith:2000-4184:PRL, Bayindir:2002-120:APL, Parazzoli:2003-107401:PRL} studies have shown the possibility
of creating novel types of microstructured materials that
demonstrate the property of negative refraction. In particular,
the composite materials created by arrays of wires and split-ring
resonators were shown to possess both negative real parts of magnetic permeability and dielectric permittivity for microwaves.
These materials are often referred to as {\em left-handed metamaterials},
{\em double-negative materials}, {\em negative-index materials}, or {\em materials with negative refraction}. Properties of such materials were first
analyzed theoretically by V. Veselago a long time ago \cite{Veselago:1967-517:UFN}, but they were demonstrated experimentally only
recently. As was shown by Veselago \cite{Veselago:1967-517:UFN}, left-handed
metamaterials possess a number of peculiar properties, including
negative refraction for interface scattering, inverse light
pressure, reverse Doppler effect, etc.

Many suggested and demonstrated applications of negative-index
metamaterials utilize unusual features of wave propagation and
scattering at the interfaces. In particular, the effect of
negative refraction can be used to realize focusing with a flat
slab, the so-called planar lens~\cite{Veselago:1967-517:UFN}; in a sharp
contrast with the well-known properties of conventional lenses with
a positive refractive index where curved surfaces are needed to
form an image. Moreover, the resolution of the negative-index flat
lens can be better than a wavelength due to the effect of
amplification of evanescent modes~\cite{Pendry:2000-3966:PRL}.

Direct numerical simulations provide the unique means for a design
of microwave and optical devices based on the negative-index
materials, however any realistic simulation should take into
account metamaterial dispersion and
losses~\cite{Fang:2003-161:APL, Cummer:2003-1503:APL, Rao:2003-67601:PRE, Feise:2005-326:PLA} as well as a nonlinear
response~\cite{Zharova:2005-1291:OE}. Such numerical simulations are often
carried out within the framework of the effective medium
approximation, when the metamaterial is characterized by the
effective dielectric permittivity and magnetic permeability. This
simplification allows for modelling of large-scale wave dynamics
using the well-known finite-difference time-domain (FDTD) numerical
methods~\cite{Taflove:2000:ComputationalElectrodynamics}.

In this paper, we discuss the main features and major difficulties
in applying the standard FDTD numerical schemes for simulating
wave scattering by wedges and interfaces of finite-extend
negative-index metamaterials, including a key issue of positioning
of a discretization grid in the numerical Yee scheme~\cite{Yee:1966-302:ITAP}
necessary to obtain consistent convergence in modeling surface
waves at an interface between conventional dielectric and
metamaterial with negative dielectric permittivity and negative
magnetic permeability. In particular, we demonstrate that, in the
framework of the continuous-medium approximation, wave scattering
on the wedge may result in a resonant excitation of surface waves
with infinitely large spatial frequencies, leading to
non-convergence of the simulation results that depend on the
discretization step.

\section{BASIC EQUATIONS}
\lsect{model}

We consider a two-dimensional problem for the propagation of
TE-polarized electromagnetic waves in the plane $(x,z$), where the
medium properties are isotropic and characterized by the
dielectric permittivity $\varepsilon$ and magnetic permeability
$\mu$. In the absence of losses, ${\cal I}m\;\varepsilon = {\rm
Im}\;\mu = 0$. The response of negative-index materials is known
to be strongly frequency dependent~\cite{Pendry:1996-4773:PRL},
however in the linear regime the wave propagation at different
wavelengths can be described independently. The stationary form of
Maxwell's equation for the complex wave envelopes is well-known
\begin{equation} \leqt{Maxwell}
      \frac{\partial H_z}{\partial x}
      - \frac{\partial H_x}{\partial z}
      = \frac{ i \omega}{c} \varepsilon(x,z) E_y,
   \quad
      \frac{\partial E_y}{\partial z}
      = - \frac{ i \omega}{c} \mu(x,z) H_x,
   \quad
      \frac{\partial E_y}{\partial x}
      = \frac{ i \omega}{c} \mu(x,z) H_z,
\end{equation}
where $H_x$, $H_z$, and $E_y$ are the components of the magnetic
and electric fields, respectively, $\omega$ is angular frequency,
and $c$ is the speed of light in vacuum.
The system of coupled equations~\reqt{Maxwell} can be reduced to a
single Helmholtz-type equation for the electric field envelope,
\begin{equation} \leqt{Helm}
   \mu(x,z) \frac{\partial}{\partial x}
        \left( \frac{1}{\mu(x,z)} \frac{\partial E_y}{\partial x} \right)
   + \mu(x,z) \frac{\partial}{\partial z}
        \left( \frac{1}{\mu(x,z)} \frac{\partial E_y}{\partial z} \right)
   + \frac{\omega^2}{c^2} n^2(x,z) E_y
   = 0,
\end{equation}
where $n^2(x,z) = \epsilon(x,z)\mu(x,z)$, and $n$ is the refractive
index of the medium.

\section{WAVE SCATTERING BY A NEGATIVE-INDEX SLAB}
          \lsect{interface}

The concept of perfect sub-wavelength imaging of a point source
through reconstitution of the evanescent waves by a flat lens has
remained highly controversial~\cite{Venema:2002-119:NAT} because it is severely
constrained by anisotropy and losses of the metamaterials.
Nevertheless, several numerical studies showed that nearly-perfect
imaging should be expected even under realistic conditions when
both dispersion and losses are taken into
account~\cite{Fang:2003-161:APL, Cummer:2003-1503:APL, Rao:2003-67601:PRE, Feise:2005-326:PLA}. In this section, we consider
the numerical simulations of the wave scattering by a slab of the
negative-index material, i.e. the problem close to that of the
perfect lens imaging, and discuss the convergence of the Yee
numerical discretization scheme.

\subsection{Geometry and discretization}

\pict{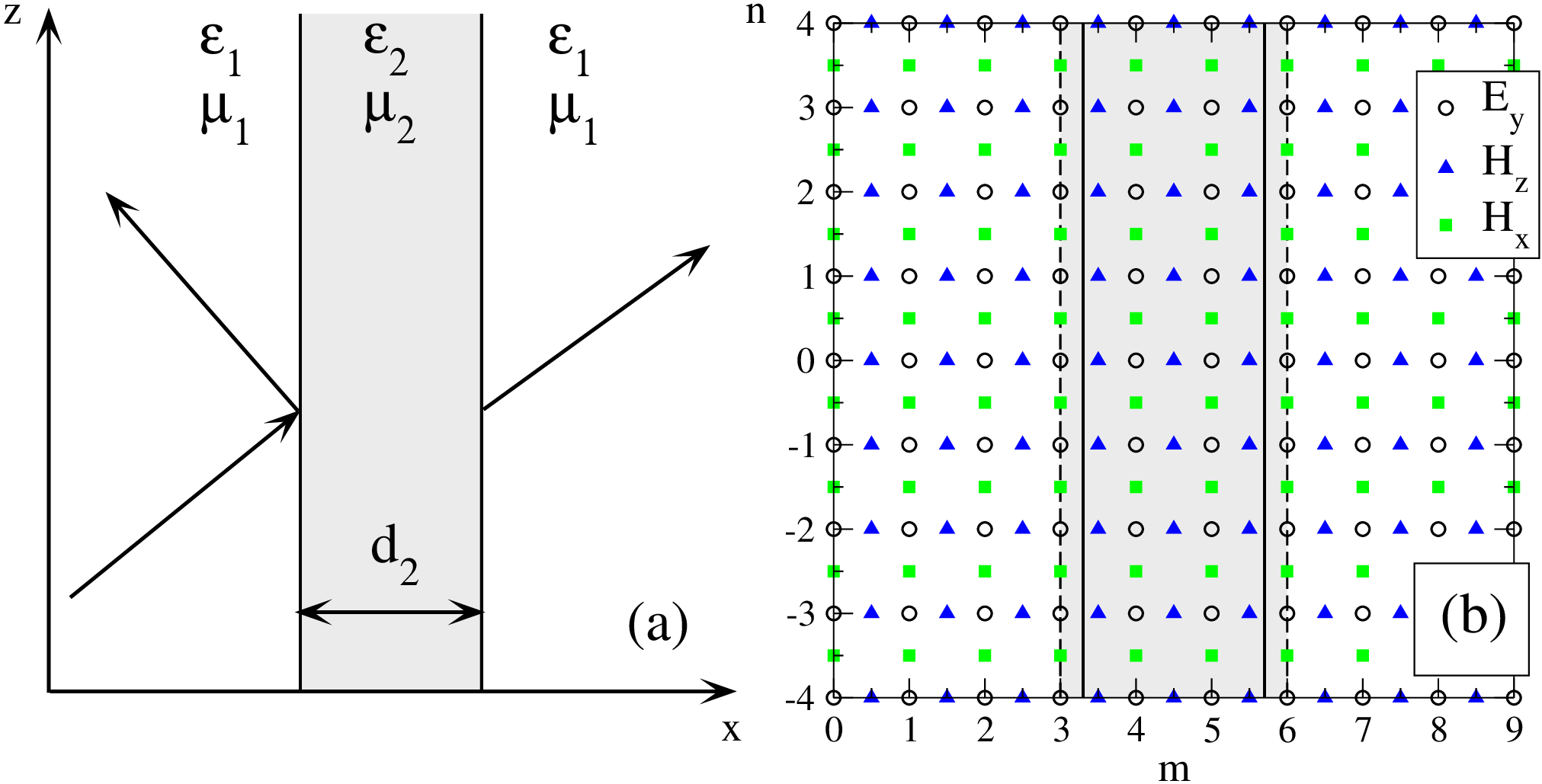}{scheme1D}{ (a)~Geometry of the scattering problem, the
slab of the thickness $d_2$ is made of a negative-index metamaterial with both $\epsilon_2$ and $\mu_2$ negative. (b)~Discretization scheme in the Yee method.}

We start our analysis by considering wave propagation through a
slab made of the negative-index material, as schematically
illustrated in Fig.~\rpict{scheme1D}, with homogeneous properties
in the $(y,z)$ plane characterized by two functions, the electric
permittivity $\varepsilon = \varepsilon(x)$ and magnetic
permeability $\mu = \mu(x)$. To solve this problem numerically, we
employ the well-known numerical Yee
method~\cite{Taflove:2000:ComputationalElectrodynamics} and perform the discretization of the electric and magnetic fields on a square grid ($x_m = h\;m$, $z_n = h\;n$) presenting the fields in the form,
\begin{equation} \leqt{Yee}
   E_y|_{m,n} = \langle E_y \rangle_{m,n}, \quad
   H_z|_{m+1/2,n} = \langle H_z \rangle|_{m+1/2,n}, \quad
   H_x|_{m,n+1/2} = \langle H_x \rangle|_{m,n+1/2}, \quad
\end{equation}
where we use the notation
$$\langle\bullet\rangle_{n,m}
 = \int_{x_{m-1/2}}^{x_{m+1/2}}\int_{z_{n-1/2}}^{z_{n+1/2}}
                       \bullet\; dx\; dz.
 $$
Then, we replace the continuum model by a closed set of the discrete equations for the field amplitudes obtained by averaging equations~\reqt{Maxwell} over the cells of discretization mesh, taking into account the continuity of the tangential field components at the
interface~\cite{Taflove:2000:ComputationalElectrodynamics},
\begin{equation} \leqt{MaxwellD}
   \begin{array}{l} {\displaystyle
      \frac{H_z|_{m+1/2,n} - H_z|_{m-1/2,n}}{h}
      - \frac{H_x|_{m,n+1/2} - H_x|_{m,n-1/2}}{h}
      = \frac{i \omega}{c} \left\langle\varepsilon\right\rangle_{m,n}
                     E_y|_{m,n} ,
   } \\*[12pt] {\displaystyle
      \frac{E_y|_{m,n+1} - E_y|_{m,n}}{h}
          \langle{\mu^{-1}}\rangle_{m,n+1/2}
      = - \frac{i \omega}{c} H_x|_{m,n+1/2} ,
   } \\*[12pt] {\displaystyle
      \frac{E_y|_{m,n} - E_y|_{m+1,n}}{h}
      = \frac{i \omega}{c}
              \left\langle\mu\right\rangle_{m+1/2,n} H_z|_{m+1/2,n} .
   } \end{array}
\end{equation}
Whereas the general form of the discrete equations~\reqt{MaxwellD}
is well known~\cite{Taflove:2000:ComputationalElectrodynamics}, we
point out a number of specific features arising in numerical
simulations of the waves scattering at the interfaces with the
negative-index media. Since the real parts of both $\varepsilon$
and $\mu$ change sign at these interfaces, the corresponding
averaged values may become small or even vanish for a certain
layer position with respect to the numerical grid. In this case,
Eqs.~\reqt{MaxwellD} may become (almost) singular, leading to poor
convergence. In this paper, we suggest that consistent convergence can be
achieved by artificially shifting the layer boundary with respect
to the grid in order to ensure that the averaged values do not vanish. This shift will not exceed $h/2$, assuring convergence as the step-size is decreased.

Because the tangential component $E_y$ of the electric field
should be continuous at the interface, is seems that a natural
choice is to align the boundary position with the grid points
$x_m$, where $E_y|_{m,n}$ is defined, and we use this configuration in
the numerical simulations presented below. However, we note that
such a selection leads to singularities for averaged values if
$\varepsilon_1=-\varepsilon_2$ or $\mu_1=-\mu_2$, which coincides with the flat-lens condition. Therefore, it is necessary to take into account losses in the metamaterial, described by nonzero imaginary parts of the complex values
$\varepsilon_2$ and $\mu_2$, or to choose a different boundary
alignment to the grid for the numerical simulations of perfect
lenses~\cite{Pendry:1996-4773:PRL}.

\pict{fig02}{spectr1D}{ (a)~Spectrum of wavenumbers $k_z$ for a
negative-index layer: exact (circles) and discrete (crosses, for
$N=512$) solutions. (b)~Absolute differences between the exact and
discrete values of $k_z$ vs. the number of points ($N$) along $x$
for the marked points 1,2,3 in~(a). Bottom: Mode profiles marked
1,2,3 in (a). The computational domain is $0 < x < 6.4$, $d_2
= 1.4$ is the width of the negative-index layer with
$\varepsilon_2 = -1.2$ and $\mu_2 = -1.5$, and $\varepsilon_1=\mu_1=1$. The wavenumber in
vacuum is normalized as $\omega/c=1$. }

\subsection{Wave spectrum and convergence of discrete solutions}

In order to illustrate the convergence of the proposed numerical
scheme, we compare the solutions of discrete and continuous
equations. We note that wave scattering from an infinite layer is
fully characterized by the properties of spatial modes, which
wavevector components along the layer ($k_z$) are conserved. These
modes have the form
\begin{equation} \leqt{mode1d}
   {\bf E}(x,z) = {\cal E}(x;\; k_z) {\rm exp}( i k_z z ), \quad
   {\bf H}(x,z) = {\cal H}(x;\; k_z) {\rm exp}( i k_z z ).
\end{equation}
Substituting Eqs.~\reqt{mode1d} into Eq.~\reqt{Maxwell} and Eq.~\reqt{MaxwellD}, we obtain a set of corresponding continuous and discrete eigenmode equations. For every $k_z$, the mode
profiles can be determined analytically, e.g., using the transfer-matrix
method~\cite{Yeh:1988:OpticalWaves}. The wave spectrum can contain
solutions corresponding to the guided modes of a negative-index
layer~\cite{Shadrivov:2003-57602:PRE}, and extended (or
propagating) modes that should also be taken into account as well, in order to describe scattering of arbitrary fields.

\pict{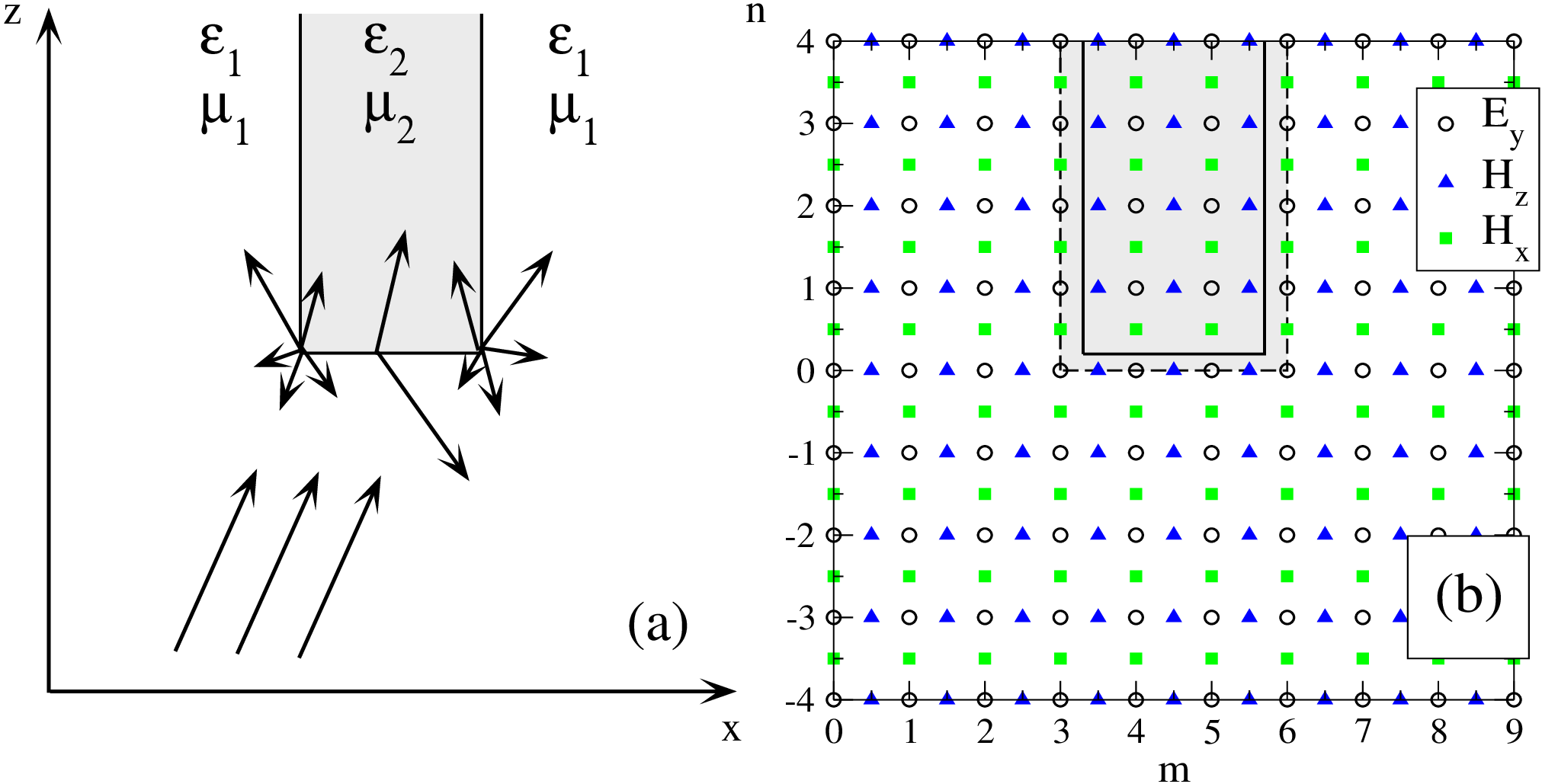}{wedge}{ (a)~Geometry of the scattering problem, the
finite-extent slab is made of a negative-index metamaterials with
both $\epsilon_2$ and $\mu_2$ negative. (b)~Discretization scheme
in the Yee method.}
\pict{fig04}{gamma}{Dependence of (a)~the cosine of the singularity parameter and (b)~its real (solid) and imaginary (dashed) parts on $\mu_2/\mu_1$ according to Eq.~\reqt{corner}.
Shading marks the region with ${\cal I}m\gamma \ne 0$.}

We solve the discrete eigenmode equations numerically for the slab
geometry with periodic boundary conditions, and compare the
spectrum of eigenvalues $k_z$ with exact solutions of the
continuous model. In Fig.~\rpict{spectr1D}(a), we show a part of the spectrum of the discrete eigenvalues (crosses), which indeed coincides with the exact values (circles). The rate of convergence can be judged from Fig.~\rpict{spectr1D}(b), where the differences between the approximate and exact solutions are shown in logarithmic scale.

\section{WAVE SCATTERING BY A WEDGE OF NEGATIVE-INDEX MATERIAL}
          \lsect{corner}

One of the fundamental problems in the theory of negative-index
metamaterials is the wave scattering by wedges~\cite{Boardman:2005-1443:JOSB},
where convergence of numerical methods can be slow due to the
appearance of singularities~\cite{Meixner:1972-442:ITAP}. In this
section, we demonstrate that the nature of such singularities has
to be taken into account when performing FDTD numerical
simulations.

\subsection{Singularity parameter}
\lsect{singularity}

\pict{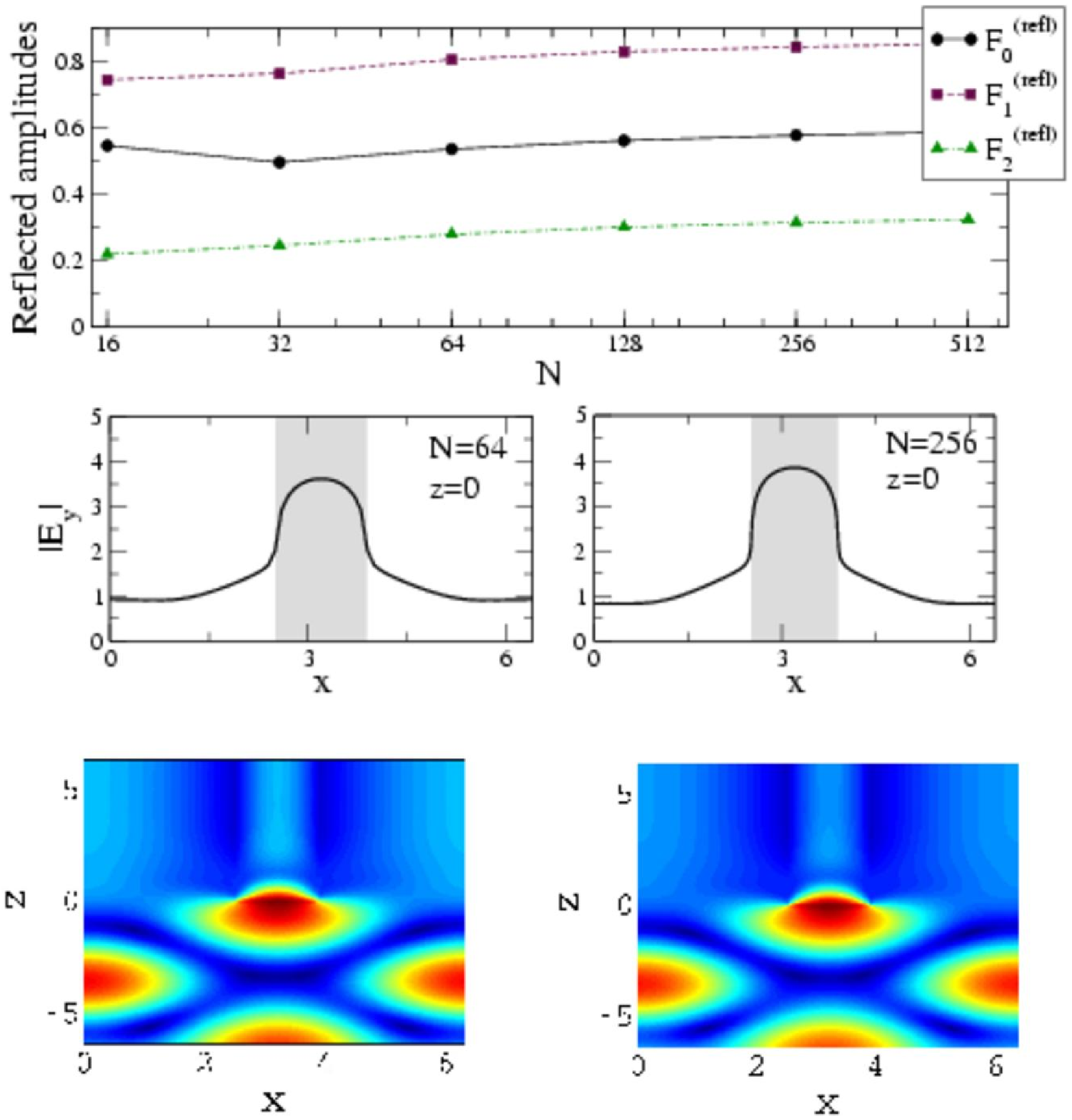}{edge35}{ (a)~Amplitudes of reflected plane waves vs.
the number of points; (b,c)~Electric field profiles for $N=64$ and
$N=256$. Parameters are the same as in Fig.~\rpict{spectr1D},
except $\mu_2 = -3.5$. }

\pict{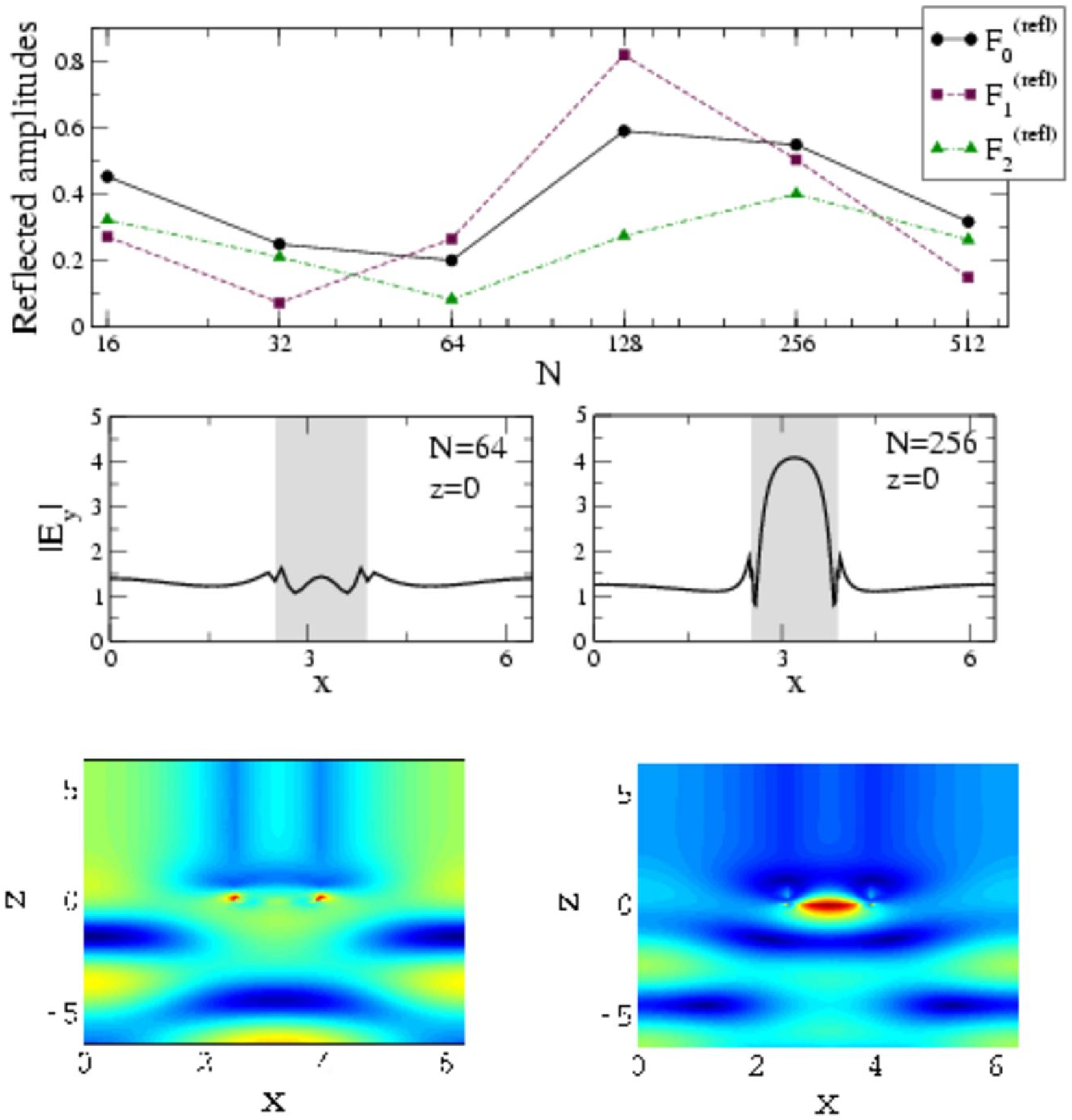}{edge15}{ (a)~Amplitudes of reflected plane waves vs.
the number of points; (b,c)~Electric field profiles for $N=64$ and
$N=256$. Parameters are the same as in Fig.~\rpict{spectr1D}. }

The behavior of the electric and magnetic fields at the wedges
between homogeneous materials characterized by different values of
$\varepsilon$ and $\mu$ was described analytically in the
pioneering paper of Meixner~\cite{Meixner:1972-442:ITAP} and
further refined in the subsequent studies (see, e.g., Ref.~\cite{Hadley:2002-1219:JLT}, and references therein). In the
case of the TE wave scattering by a negative-index wedge, as
schematically illustrated in Fig.~\rpict{wedge}, the amplitudes of
magnetic fields exhibit singular behavior at the wedge of the
order of $\rho^{\gamma-1}$, where $\rho$ is the distance from the
wedge. For a  $\pi/2$ wedge angle, corresponding to a corner of a
rectangular slab, the coefficient $\gamma$ is found
as~\cite{Meixner:1972-442:ITAP}
\begin{equation} \leqt{corner}
   \frac{\mu_1 - \mu_2}{\mu_1 + \mu_2}
   = \pm 2 {\rm cos}\left( \gamma \pi / 2 \right) ,
\end{equation}
where $\mu_1$ and $\mu_2$ are magnetic permeabilities of the two
neighboring media. In the case of conventional dielectric or
magnetic media with $\mu_j>0$, Eq.~\reqt{corner} has solutions
with real $\gamma$. However, when $\mu$ changes its sign at the
interface with a negative-index medium, the coefficient $\gamma$
becomes {\em complex} playing role of a singularity parameter. This
happens when
\begin{equation} \leqt{muRange}
   - 3 \mu_1 < \mu_2 < -\mu_1 / 3 ,
\end{equation}
so that $|\mu_1-\mu_2|/|\mu_1+\mu_2| > 2$, see Fig.~\rpict{gamma}.

For real $\gamma$ (taking the solution with $0<\gamma<1$, according
to Ref.~\cite{Meixner:1972-442:ITAP}), the field amplitudes decay
monotonously away from the corner. In this case, numerical
simulations may be based on the simplest discretization, although
the convergence rate can be improved by taking into account the
singular behavior in the discrete equations.

However, the case of complex $\gamma$ corresponds to the fields
that {\em oscillate infinitely fast} near the corner because
$$\rho^{\gamma-1} = \rho^{{\cal R}e(\gamma-1)}
               \exp\left[i\, {\cal I}m(\gamma)\,\log(\rho) \right].
               $$
The second multiplier indicates the excitation of infinitely large
spatial harmonics, however such a situation is unphysical because the
effective-medium approximation of the negative-index materials is valid for slowly varying fields only. Therefore, in numerical simulations it is necessary to take into account the physical effects that suppress such oscillations, in particular we discuss the effect of losses in Sec.~\rsect{losses} below.

\subsection{Numerical results}

We now analyze convergence of the numerical finite-difference
solutions for the problem of wave scattering by a finite-extent
negative-index slab. We align the boundaries of the negative-index
domain with the grid, as shown schematically in Fig.~\rpict{scheme1D}(b). Since the electric field components $E_y$ are continuous at the interfaces, it is possible to obtain the discrete equations that have the form of Eqs.~\reqt{MaxwellD}.

\pict{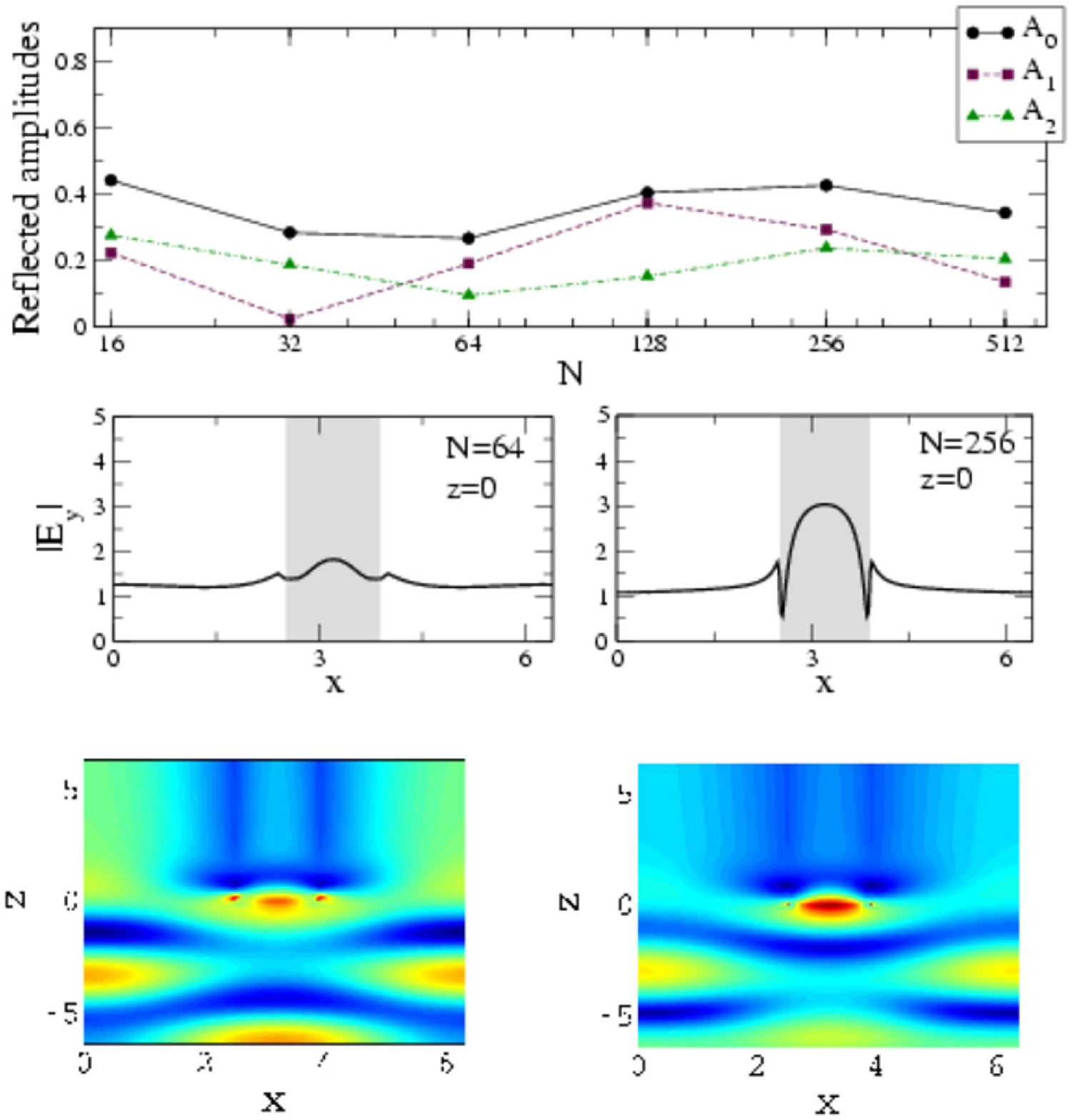}{edge15loss}{ Same as in Fig.~\rpict{edge15} but for $\mu = -1.5 + i\;10^{-1}$. }

In order to construct the full solutions for  scattering problem,
we decompose the field into a set of eigenmodes of the
negative-index layer ($z>0$) and free space ($z<0$). More
specifically, for $z>0$ we have
\begin{equation}\leqt{blochDiscr}
   {\bf E|_{m,n}} = 
   \sum_j A_j {\cal E}|_{m,n}(k_z^{(j)}) {\rm exp}( i k_z^{(j)} z_n ), \quad
   {\bf H|_{m,n}} = 
   \sum_j A_j {\cal H}|_{m,n}(k_z^{(j)}) {\rm exp}( i k_z^{(j)} z_n ),
\end{equation}
where $j$ is the number of eigenmodes. Here the summation is
performed over {\em the propagating modes} (${\cal I}m k_z = 0$) which
transport energy away from the interface, and {\em evanescently
decaying modes} with ${\cal I}m k_z < 0$. In free space at
$z<0$, the field is composed of incident and reflected plane
waves,
\begin{equation}\leqt{planeDiscr}
   E_y|_{m,n} = \sum_{j=-N/2+1}^{N/2} F_j^{\rm (in)} 
                \exp( i 2 \pi j m /  N + i K^{(j)} z_n)
              + \sum_{j=-N/2+1}^{N/2} F_j^{\rm (refl)} 
                \exp( i 2 \pi j m / N - i K^{(j)} z_n ) .
\end{equation}
Here $N$ is the number of points in the $x$ direction, and the
discrete wavenumber is $K^{(j)} = \pm {\rm sin}^{-1} \left[
h^2\omega^2/4c^2-{\rm sin}^2 (j \pi/N)\right]^{1/2} 2/h$ \cite{Taflove:2000:ComputationalElectrodynamics}. The sign of $K^{(j)}$ is chosen with a proper wave asymptotic behavior, i.e., we choose ${\cal R}e (K^{(j)}) > 0$, if $K^{(j)}$ is real, and  ${\cal I}m (K^{(j)}) > 0$ if $K^{(j)}$ is complex. The magnetic field at $z<0$ is found from Eq.~\reqt{MaxwellD}, with homogeneous parameters $\langle \varepsilon \rangle = \varepsilon_1$, $\langle \mu \rangle = \mu_1$, $\langle \mu^{-1} \rangle = \mu_1^{-1}$.  Then, we substitute Eqs.~\reqt{blochDiscr},~\reqt{planeDiscr} into the first of Eq.~\reqt{MaxwellD}, and using the condition of the continuity of the electric field, obtain a set of equations for all $m=1,\ldots,N$ that are used to calculate the amplitudes $F_j^{(refl)}$ and $A_j$ of the reflected and transmitted waves,
\begin{equation}
   \begin{array}{l} {\displaystyle
      \sum_j A_j {\cal E}_y|_{m,0}(k_z^{(j)})
      = \sum_{j=-N/2+1}^{N/2} (F_j^{\rm (in)}+ F_j^{\rm (refl)})
                    \exp( i 2 \pi j m / N)
   } \\*[12pt] {\displaystyle
      \sum_j A_j \left[ {\cal H}_x|_{m,-1/2}(k_z^{(j)})
                        + {\cal H}_x|_{m,1/2}(k_z^{(j)}) \right]
   } \\*[12pt] {\displaystyle \qquad
      = \frac{- 2 c}{\omega \mu_1 h} 
        \sum_{j=-N/2+1}^{N/2} 
        (F_j^{\rm (in)}-F_j^{\rm (refl)})
         \sin(K^{(j)} h) \exp( i 2 \pi j m / N) .
   } \end{array}
\end{equation}
These equations are solved using the standard linear algebra package.

\pict{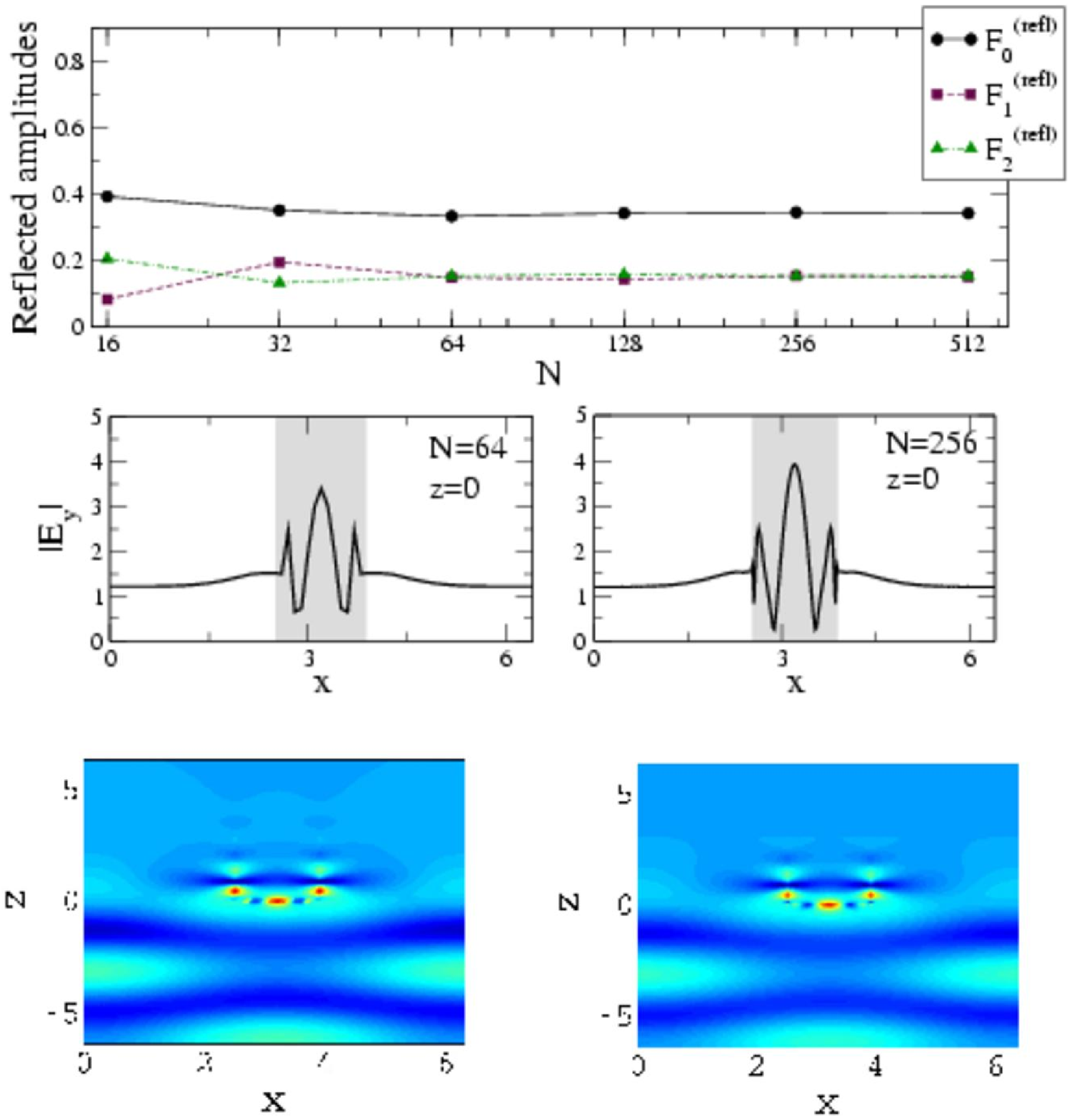}{edge101Loss}{ Same as in Fig.~\rpict{edge15} but for $\mu = -1.01 + i\;10^{-2}$. }

We now present results of our numerical simulations for the
scattering of normally incident plane waves, with $F_0^{(in)} =
1$, $F_{j \ne 0}^{(in)} = 0$. First, we consider scattering by a negative-index slab with $\mu_2 < -3$. In this case, we observe a steady convergence of
numerical solutions, as shown in Fig.~\rpict{edge35}. This demonstrates that even simplest finite-difference numerical schemes can be successfully employed to model the scattering process when the sinularity parameter $\gamma$ is real, is in a full agreement with earlier studies of wave scattering at dielectric wedges~\cite{Hadley:2002-1219:JLT}.

However, the situation changes dramatically when $\gamma$ is complex, i.e. for $\mu_2 = -1.5$.
According to the analytical solution, in this case the magnetic
field should oscillate infinitely fast in the vicinity of the
corner, corresponding to excitation of infinitely large spatial
harmonics. However, such behavior cannot be described by discrete
equations, and we find that in this regime {\em solutions of
finite-difference equation do not converge}, as demonstrated in
Fig.~\rpict{edge15}. 

\subsection{Effects of losses} \lsect{losses}

The analytical description of the edge singularities discussed
above is only valid for lossless media, i.e. when all
$\varepsilon$ and $\mu$ are real. However, the negative-index
metamaterials always have non-vanishing losses, and we have
studied whether this important physical effect can regularize the field
oscillations at the corner. However, our results demonstrate that
even substantial losses may not be sufficient enough to suppress such
oscillations, as presented in the example of
Fig.~\rpict{edge15loss}.

\subsection{Singularities and perfect lenses} 

Finally, we consider the problem of wave scattering from the corners of perfect lenses, where ${\cal R}e \left(\varepsilon_2 \right) \simeq -1$ and ${\cal R}e \left( \mu_2 \right) \simeq - 1$ (we take $\varepsilon_1=\mu_1=1$). This is a special case, where the type of singularity becomes indefinite if
losses are neglected. We find that introducing {\em sufficiently
large losses} does indeed regularize the field
oscillations at the corners , leading to convergence of numerical simulations, as demonstrated for the example of Fig.~\rpict{edge101Loss}. However, this only occurs when the value of losses exceeds a certain
threshold; if the losses are too weak then non-convergent behavior is again observed, as shown in Fig.~\rpict{edge101LossLow}. The threshold value of losses for achieving convergence of the numerical scheme is increased for larger 
$|{\cal R}e\, \left( \mu_2 \right) +1|$. We note that this is completely different from the temporal dynamics at an infinitely extended slab, where convergence to a steady state is eventually achieved with arbitrarily small losses~\cite{Rao:2003-67601:PRE}.

\pict{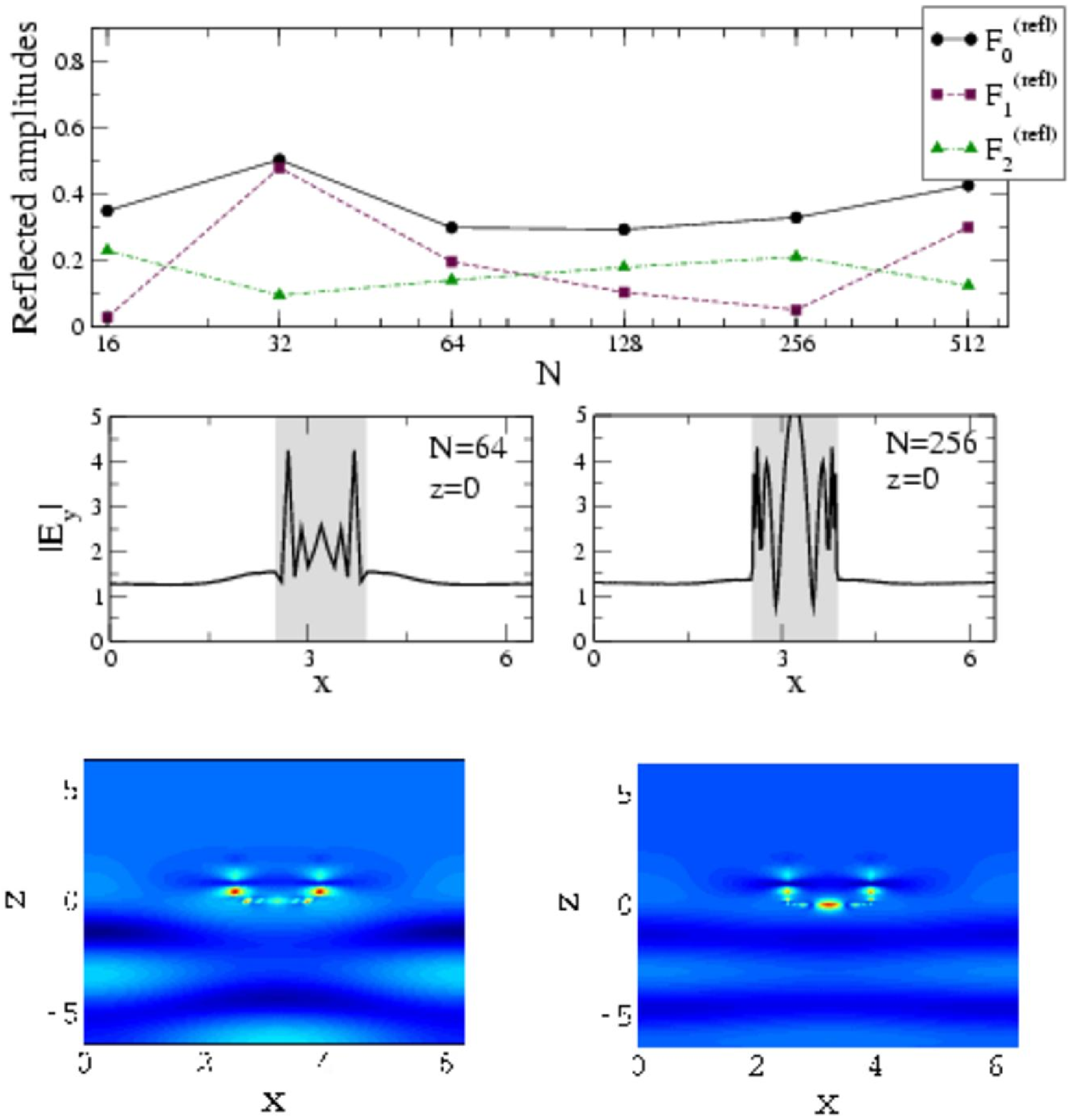}{edge101LossLow}{
Same as in Fig.~\rpict{edge15} but for $\mu = -1.01 + i\;10^{-3}$.
}

\section{CONCLUSIONS}

We have discussed a number of specific features manifested in numerical
simulations of wedges and interfaces of metamaterials, i.e.
composite materials with negative dielectric permittivity and negative
magnetic permeability. We have demonstrated that a numerical
discretization grid in the Yee method may have a dramatic effect
on the convergence in numerical modelling of surface waves at
interfaces and wedges. In the framework of the continuous-medium
approximation, wave scattering on the wedge may result in a
resonant excitation of surface waves with infinitely large spatial
frequencies, leading to non-convergence of the numerical
simulation results that depend strongly on the value of the discretization
step. We find that sufficiently high losses may suppress oscillations and allow to obtain converging solutions to the scattering problem, however in the case of smaller losses it may be necessary to take into account the meta-material properties beyond the effective-medium approximation, such as the effect of spatial dispersion.

\section*{ACKNOWLEDGEMENTS}

The authors thank Alexander~Zharov and Pavel~Belov for useful
discussions and suggestions. This work has been supported by the
Australian Research Council.


\begin{thebibliography}{10}

\bibitem{Pendry:1996-4773:PRL}
J.~B. Pendry, A.~J. Holden, W.~J. Stewart, and I. Youngs, ``Extremely low
  frequency plasmons in metallic mesostructures,'' Phys. Rev. Lett. {\bf 76,}
  4773--4776 (1996).

\bibitem{Pendry:1999-2075:ITMT}
J.~B. Pendry, A.~J. Holden, D.~J. Robbins, and W.~J. Stewart, ``Magnetism from
  conductors and enhanced nonlinear phenomena,'' IEEE Trans. Microw. Theory
  Tech. {\bf 47,} 2075--2084 (1999).

\bibitem{Markos:2002-36622:PRE}
P. Markos and C.~M. Soukoulis, ``Numerical studies of left-handed materials and
  arrays of split ring resonators,'' Phys. Rev. E {\bf 65,} 036622--8 (2002).

\bibitem{Markos:2002-33401:PRB}
P. Markos and C.~M. Soukoulis, ``Transmission studies of left-handed
  materials,'' Phys. Rev. B {\bf 65,} 033401--4 (2002).

\bibitem{Smith:2000-4184:PRL}
D.~R. Smith, W.~J. Padilla, D.~C. Vier, S.~C. Nemat~Nasser, and S. Schultz,
  ``Composite medium with simultaneously negative permeability and
  permittivity,'' Phys. Rev. Lett. {\bf 84,} 4184--4187 (2000).

\bibitem{Bayindir:2002-120:APL}
M. Bayindir, K. Aydin, E. Ozbay, P. Markos, and C.~M. Soukoulis, ``Transmission
  properties of composite metamaterials in free space,'' Appl. Phys. Lett. {\bf
  81,} 120--122 (2002).

\bibitem{Parazzoli:2003-107401:PRL}
C.~G. Parazzoli, R.~B. Greegor, K. Li, B.~E.~C. Koltenbah, and M. Tanielian,
  ``Experimental verification and simulation of negative index of refraction
  using Snell's law,'' Phys. Rev. Lett. {\bf 90,} 107401--4 (2003).

\bibitem{Veselago:1967-517:UFN}
V.~G. Veselago, ``The electrodynamics of substances with simultaneously
  negative values of $\varepsilon$ and $\mu$,'' Usp. Fiz. Nauk {\bf 92,}
  517--526 (1967) (in Russian) [English translation: Phys. Usp. {\bf 10,} 509--514 (1968)].

\bibitem{Pendry:2000-3966:PRL}
J.~B. Pendry, ``Negative refraction makes a perfect lens,'' Phys. Rev. Lett.
  {\bf 85,} 3966--3969 (2000).

\bibitem{Fang:2003-161:APL}
N. Fang and X. Zhang, ``Imaging properties of a metamaterial superlens,'' Appl.
  Phys. Lett. {\bf 82,} 161--163 (2003).

\bibitem{Cummer:2003-1503:APL}
S.~A. Cummer, ``Simulated causal subwavelength focusing by a negative
  refractive index slab,'' Appl. Phys. Lett. {\bf 82,} 1503--1505 (2003).

\bibitem{Rao:2003-67601:PRE}
X.~S. Rao and C.~K. Ong, ``Subwavelength imaging by a left-handed material
  superlens,'' Phys. Rev. E {\bf 68,} 67601--3 (2003).

\bibitem{Feise:2005-326:PLA}
M.~W. Feise and Yu.~S. Kivshar, ``Sub-wavelength imaging with a left-handed
  material flat lens,'' Phys. Lett. A {\bf 334,} 326--330 (2005).

\bibitem{Zharova:2005-1291:OE}
N.~A. Zharova, I.~V. Shadrivov, A.~A. Zharov, and Yu.~S. Kivshar, ``Nonlinear
  transmission and spatiotemporal solitons in metamaterials with negative
  refraction,'' Optics Express {\bf 13,} 1291--1298 (2005).

\bibitem{Taflove:2000:ComputationalElectrodynamics}
A. Taflove and S.~C. Hagness, {\em {Computational Electrodynamics: The
  Finite-Difference Time-Domain Method}}, 2nd  ed. (Artech House, Norwood,
  2000).

\bibitem{Yee:1966-302:ITAP}
K.~S. Yee, ``Numerical solution of initial boundary value problems involving
  Maxwells equations in isotropic media,'' IEEE Trans. Antennas Propag. {\bf
  AP14,} 302 (1966).

\bibitem{Venema:2002-119:NAT}
L. Venema, ``Negative refraction: A lens less ordinary,'' Nature {\bf 420,}
  119--120 (2002).

\bibitem{Yeh:1988:OpticalWaves}
P. Yeh, {\em {Optical Waves in Layered Media}} (John Wiley \& Sons, New York,
  1988).

\bibitem{Shadrivov:2003-57602:PRE}
I.~V. Shadrivov, A.~A. Sukhorukov, and Yu.~S. Kivshar, ``Guided modes in
  negative-refractive-index waveguides,'' Phys. Rev. E {\bf 67,} 057602--4
  (2003).

\bibitem{Boardman:2005-1443:JOSB}
A.~D. Boardman, L. Velasco, N. King, and Y. Rapoport, ``Ultra-narrow bright
  spatial solitons interacting with left-handed surfaces,'' J. Opt. Soc. Am. B
  {\bf 22,} 1443--1452 (2005).

\bibitem{Meixner:1972-442:ITAP}
J. Meixner, ``The behavior of electromagnetic fields at edges,'' IEEE Trans.
  Antennas Propag. {\bf AP-20,} 442--446 (1972).

\bibitem{Hadley:2002-1219:JLT}
G.~R. Hadley, ``High-accuracy finite-difference equations for dielectric
  waveguide analysis II: dielectric corners,'' J. Lightwave Technol. {\bf 20,}
  1219--1231 (2002).

\end{thebibliography}
\end{document}